\documentstyle[epsfig,12pt,a4p,subfigure]{article}

\newcommand{\pipipipi}{\mbox{$\pi^+\pi^-\pi^+\pi^-$ }}

\newcommand{\pipi}{\mbox{$\pi^{+}\pi^{-}$} }

\newcommand{\fmeson}{\mbox{$f_{2}(1270)$} }

\begin{document}
\begin{titlepage}
\def\footnoterule{\hrule width 1.0\columnwidth}
\bigskip
\bigskip
\begin{center}{\Large  {\bf
A search for non-$q \overline q$ mesons at
the CERN Omega Spectrometer}
}\end{center}
\bigskip
\bigskip
\begin{center}{
A.\thinspace Kirk
}\end{center}
\bigskip
\bigskip
\begin{tabbing}
aba \=   \kill
\> \small
School of Physics and Astronomy, University of Birmingham, Birmingham, U.K. \\
\end{tabbing}
\begin{center}{\bf {{\bf Abstract}}}\end{center}

{
The non-Abelian nature of QCD suggests that
particles that have a gluon constituent, such as glueballs or
hybrids, should exist.
Experiments WA76, WA91 and WA102
have performed a dedicated search for these states
in central production using
the CERN Omega Spectrometer.
Several
non-$q \overline q $ candidates have been observed. This paper
presents
a study of central meson production as a function of the
difference in transverse momentum ($dP_T$)
of the exchanged particles
which shows that undisputed $q \overline q$ mesons
are suppressed at small $dP_T$ whereas the glueball candidates
are enhanced.
}
\bigskip
\bigskip
\begin{center}
{\bf Dedicated to the memory of my colleague and friend Yuri Prokoshkin}
\end{center}
\bigskip
\bigskip\begin{center}{{Submitted to Yadernaya Fizika}}
\end{center}
\end{titlepage}
\setcounter{page}{2}

\section{Introduction}
\par
The present understanding of strong interactions is that they are
described
by Quantum ChromoDynamics (QCD). This non-Abelian field theory
not only describes how quarks and antiquarks interact, but also
predicts that the gluons which are the quanta of the field will themselves
interact to form mesons.
If the object formed is composed entirely of valence gluons the meson
is called a glueball, however if it is composed of a mixture of
valence quarks, antiquarks
and gluons (i.e. $ q \overline q g$ ) it is called a hybrid.
In addition, $ q \overline q q \overline q $ states are also predicted.
\par
The best estimate for the masses of glueballs comes from
lattice gauge theory calculations
\cite{re:lgt}
which show that the lightest glueball has $J^{PC}$~=~$0^{++}$ and that
\begin{center}
$m(2^{++})/m(0^{++}) =1.5 $
\end{center}
and  depending on the extrapolation used from the lattice parameters to
mass scale that
\begin{center}
$m(0^{++}) =(1500-1750) $MeV.
\end{center}
The mass of the $0^{-+}$ glueball is predicted to be similar to that of the
$2^{++}$ glueball whilst glueballs with other quantum numbers
are predicted to be higher in mass.
\par
The flux tube model has been used to calculate the masses of the lowest
lying hybrid states and recent predictions
\cite{re:ISGURBNL}
are that
\begin{center}
$m(1^{--},0^{-+},1^{-+},2^{-+}) \approx 1900$~MeV.
\end{center}
\par
Hence, these non-$q \overline q$ states
are predicted to be in the same mass range
as the normal  $q \overline q$ nonet members
and hence we need a method of identifying them.
\par
The following have been suggested as possible ways to identify
gluonic states.
\begin{itemize}
\item
To look for "oddballs": States with $J^{PC}$ quantum numbers not allowed
for normal $ q \overline q$ states. For example $J^{PC} = 1^{-+}$.
\item
However the lightest non-$q \overline q$ states
are predicted to have the same quantum numbers as $q \overline q$ states.
Therefore we need
to look for extra states, that is states that have quantum numbers
of already completed nonets and that have masses which are sufficiently
low that they are unlikely to be members of the radially excited nonets and
hence they can not be described as being pure $ q \overline q $ states.
\item
If extra states are found then in order to isolate which state is the
likely non-$q \overline q$ state we can
\begin{enumerate}
\item[a)]
Look for states with unusual branching ratios.
\item[b)]
Look for states preferentially produced in gluon rich processes.
These processes are described below.
\end{enumerate}
\end{itemize}
\par
\begin{figure}[tp]
\begin{center}
\epsfig{file=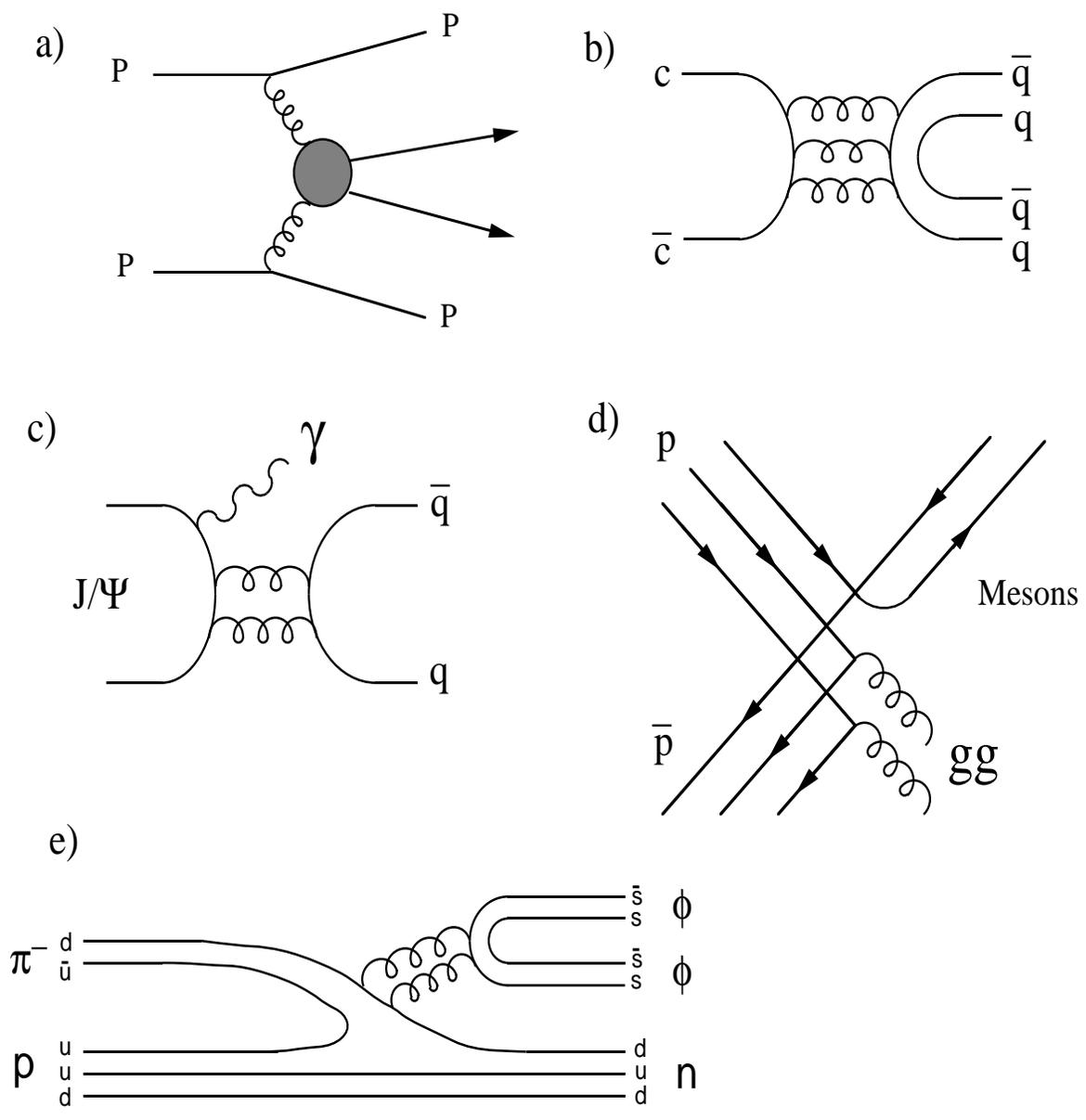,height=16cm,width=16cm}
\caption{Gluon rich channels.
Dynamical configurations that have been used to study light hadron spectroscopy
in a search for glueball states.}
\label{fi:dysum}
\end{center}
\end{figure}

Fig.~\ref{fi:dysum}
summarises several dynamical configurations which have been suggested
as possible
sources of gluonium and where experiments have been performed.
\begin{enumerate}
\item
Pomeron-Pomeron scattering is shown in fig.~\ref{fi:dysum}a).
The Pomeron is
an object which can be described as
a multi-gluon state, and
is thought to be
responsible for the large cross sections of diffractive reactions.
Consequently
Double Pomeron Exchange (DPE) is considered to be
a possible source of glueballs.
\item
The $J/\psi$ decay is believed to be
a highly glue rich channel either via the hadronic decay
shown in fig.~\ref{fi:dysum}b), or via the radiative decay shown in
fig.~\ref{fi:dysum}c).
\item
Figure~\ref{fi:dysum}d)
shows proton-antiproton annihilation; the annihilation region
of quarks and antiquarks is a source of gluons where glueballs and hybrids
could be produced.
\item
Special hadronic reactions, an example of which is shown in
fig.~\ref{fi:dysum}e)
where the $\phi\phi$ system is thought to be
produced via an intermediate state containing
gluons. Reactions of this kind which have disconnected quark lines
are said to be OZI violating
\cite{re:OZI}.
\end{enumerate}
\par
The first reaction is the one studied
by experiments WA76, WA91 and WA102 at the Omega spectrometer.
In this paper the  status of these experiments is reviewed and the possibility
of a glueball-$q \overline q$ filter in central production is discussed.
\section{The Omega Central production experiments}
\par
There is considerable current interest in trying to isolate the lightest
glueball.
Several experiments have been performed using glue-rich
production mechanisms.
One such mechanism is Double Pomeron Exchange (DPE) where the Pomeron
is thought to be a multi-gluonic object.
Consequently it has been
anticipated that production of
glueballs may be especially favoured in this process~\cite{closerev}.
\par
The Omega central production experiments
(WA76, WA91 and WA102) are
designed to study exclusive final states
formed in the reaction
\begin{center}
pp$\longrightarrow$p$_{f}X^{0}$p$_s$,
\end{center}
where the subscripts $f$ and $s$ refer to the fastest and slowest
particles in the laboratory frame respectively and $X^0$ represents
the central system. Such reactions are expected to
be mediated by double exchange processes
where both Pomeron and Reggeon exchange can occur.
\par
The trigger was designed to enhance double exchange
processes with respect to single exchange and elastic processes.
Details of the trigger conditions, the data
processing and event selection
have been given in previous publications~\cite{re:expt}.
\section{The possibility of a Glueball-$q \overline q$ filter in central
production }
\par
The experiments have been
performed at incident beam momenta of 85, 300 and 450 GeV/c, corresponding to
centre-of-mass energies of
$\sqrt{s} = 12.7$, 23.8 and 28~GeV.
Theoretical
predictions \cite{pred} of the evolution of
the different exchange mechanisms with centre
of mass energy, $\sqrt{s}$, suggest that
\begin{center}
$\sigma$(RR) $\sim s^{-1}$,\\
$\sigma$(RP) $\sim s^{-0.5}$,\\
$\sigma$(PP) $\sim$ constant,
\end{center}
where RR, RP and PP refer to Reggeon-Reggeon, Reggeon-Pomeron and
Pomeron-Pomeron
exchange respectively. Hence we expect Double Pomeron Exchange
(DPE) to be more significant at high energies, whereas the Reggeon-Reggeon and
Reggeon-Pomeron mechanisms will be of decreasing importance.
The decrease of the non-DPE cross section with energy can be inferred
from data
taken by the WA76 collaboration using pp interactions at $\sqrt{s}$ of 12.7 GeV
and 23.8 GeV \cite{wa76}.
The \pipi mass spectra for the two cases show that
the signal-to-background ratio for the $\rho^0$(770)
is much lower at high energy, and the WA76 collaboration report
that the ratio of the $\rho^0$(770) cross sections at 23.8 GeV and 12.7 GeV
is 0.44~$\pm$~0.07.
Since isospin 1 states such as the $\rho^0$(770) cannot be produced by DPE,
the decrease
of the $\rho^{0}(770)$ signal at high $\sqrt{s}$
is consistent with DPE becoming
relatively more important with increasing energy with respect to other
exchange processes.
\par
\begin{figure}[htp]
\begin{center}
\mbox{\begin{tabular}{l}
\subfigure{\mbox{\epsfig{file=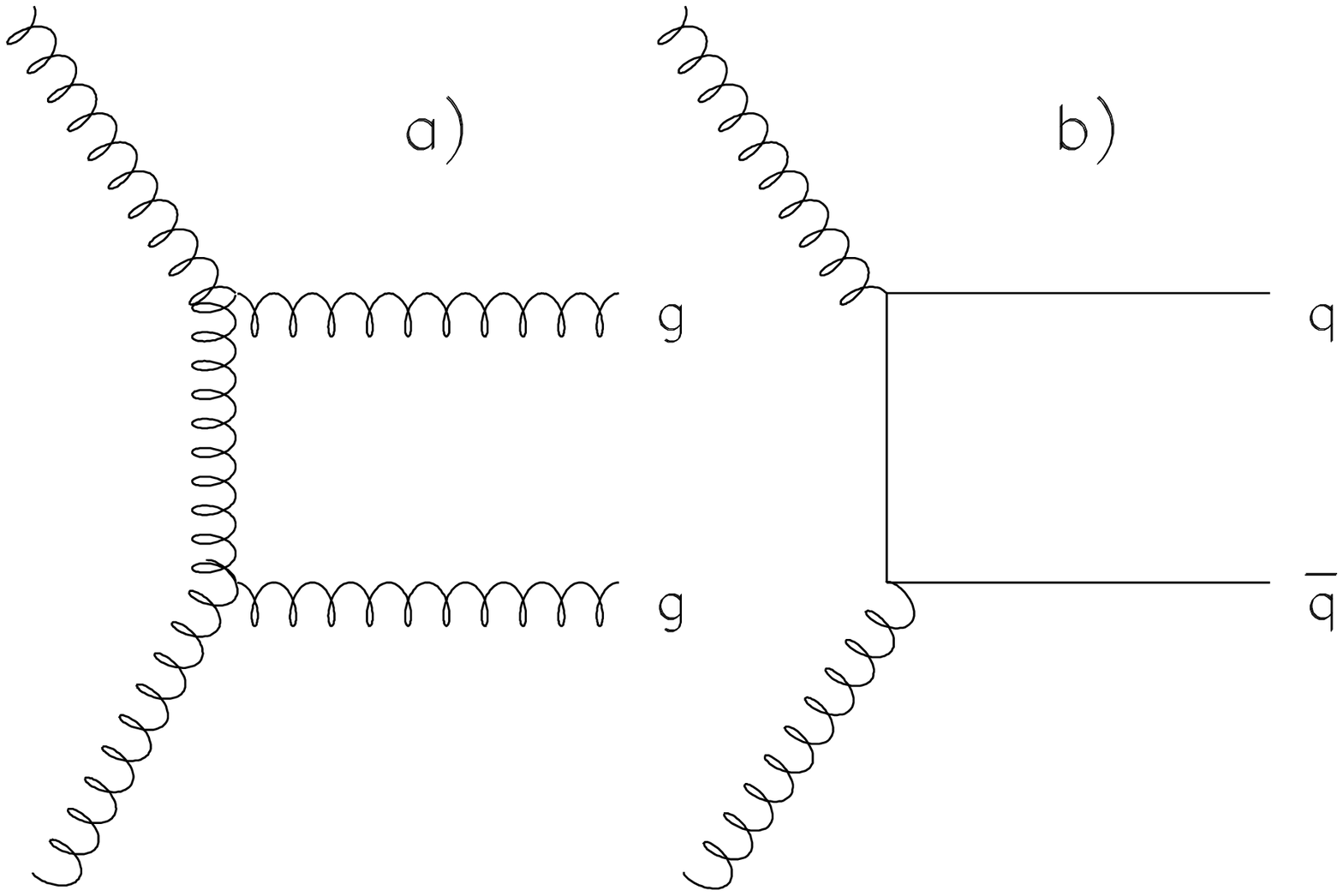,height=6cm,width=14.0cm,
bbllx=0pt,bblly=0pt,bburx=550pt,bbury=300pt}}}  \\
\subfigure{\mbox{\epsfig{file=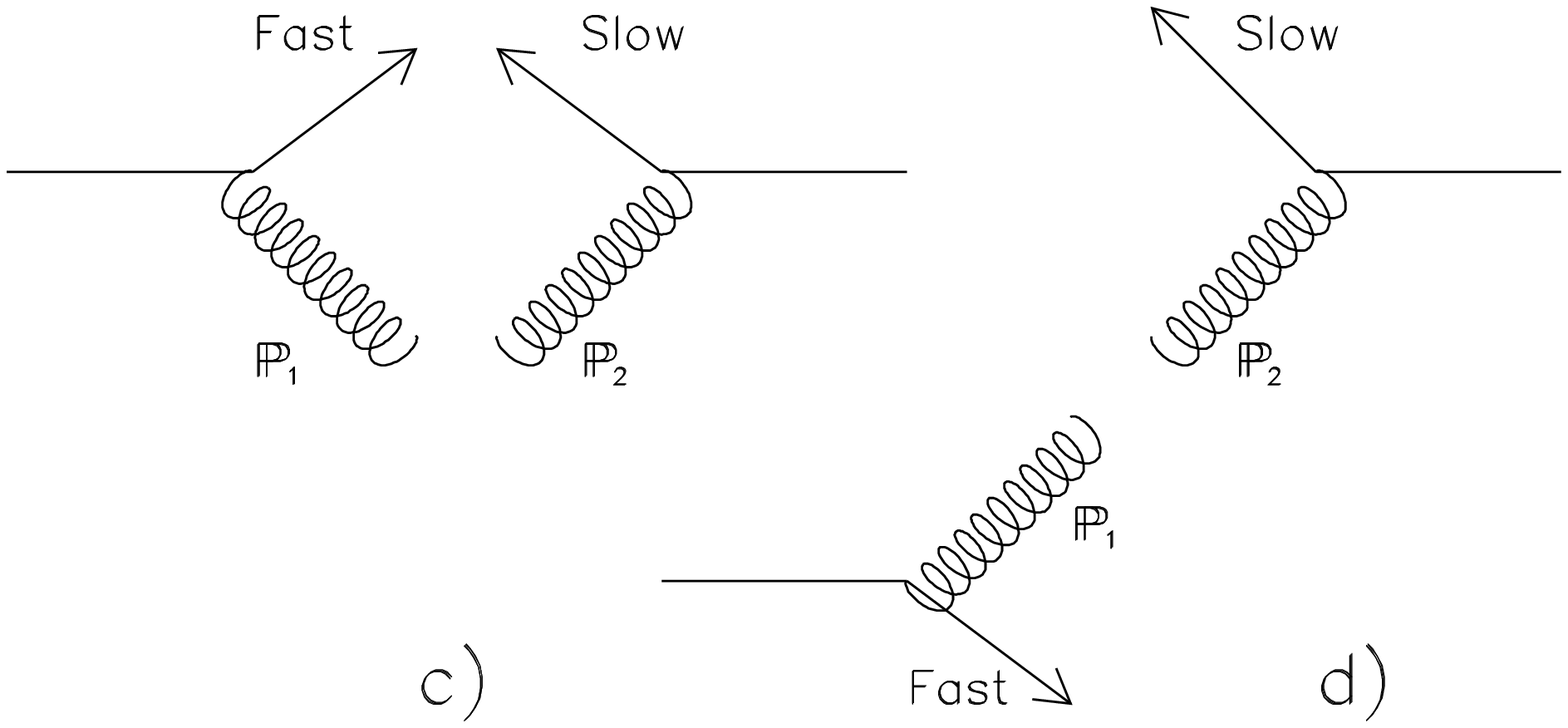,width=14.0cm}}}  \\
\end{tabular}}
\end{center}
\caption{Schematic diagrams
of the coupling of the exchange particles into the final state meson
for a) gluon exchange and b) quark exchange.
Schematic diagrams
in the CM for c) LL and d) LR triggers.}
\label{fi:feyn1}
\end{figure}

However,
even in the case of pure DPE
the exchanged particles still have to couple to a final state meson.
The coupling of the two exchanged particles can either be by gluon exchange
or quark exchange. Assuming the Pomeron
is a colour singlet gluonic system if
a gluon is exchanged then a gluonic state is produced, whereas if a
quark is exchanged then a $q \overline q $ state is produced
(see figures~\ref{fi:feyn1}a) and b) respectively).
It has been suggested recently~\cite{closeak} that
for small differences in transverse momentum between the two
exchanged particles
an enhancement in the production of glueballs
relative to $q \overline q$ states may occur.
\par
Recently the WA91 collaboration has published a paper~\cite{wa91corr}
showing that the observed centrally produced resonances depend on the
angle between the outgoing slow and fast protons.
In order to describe the data in terms of a physical model,
Close and Kirk~\cite{closeak}
have proposed that the data be analysed
in terms of the difference in transverse momentum
between the particles exchanged from the
fast and slow vertices.
\par
The trigger is described in detail in ref.~\cite{wa91corr}.
In brief, the trigger separates the data into two categories.
One where the slow and
fast particles are on the same side of the beam, {\em i.e.} a small
azimuthal angle
between the outgoing protons (classified as LL)
and one where the slow and
fast particles are on the opposite side of the beam, {\em i.e.} the azimuthal
angle between the outgoing protons is near to 180~degrees,
(classified as LR).
\par
In ref.~\cite{wa91corr} it was shown that the centrally produced
resonances depended on the trigger type i.e. the resonances
observed in the LL  trigger were different to those observed in the LR trigger.
This difference was not due to any acceptance or trigger bias but appeared to
be related to the
angle between the outgoing protons.
Figures~\ref{fi:feyn1}c) and d) show schematic representations of the
LL and LR triggers respectively in the centre of mass of the beam plus target
where the longitudinal (x) axis is defined to be along the beam direction.
In the case of the LL trigger (figure~\ref{fi:feyn1}c)) the transverse
momentum vector of each exchanged particle has the same sign whereas for
the LR trigger (figure~\ref{fi:feyn1}d))
they have the opposite sign. Hence the
difference in the transverse momentum vectors
of the two exchanged particles is greater in the LR
trigger than
in the LL trigger.
The difference in the transverse momentum vectors ($dP_T$) is defined to be
\begin{center}
$dP_T$ = $\sqrt{(P_{y1} - P_{y2})^2 + (P_{z1} - P_{z2})^2}$
\end{center}
where
$Py_i$, $Pz_i$ are the y and z components of the momentum
of the $ith$ exchanged particle in the pp centre of mass system.
\par
\begin{figure}[htp]
\begin{center}
\epsfig{file=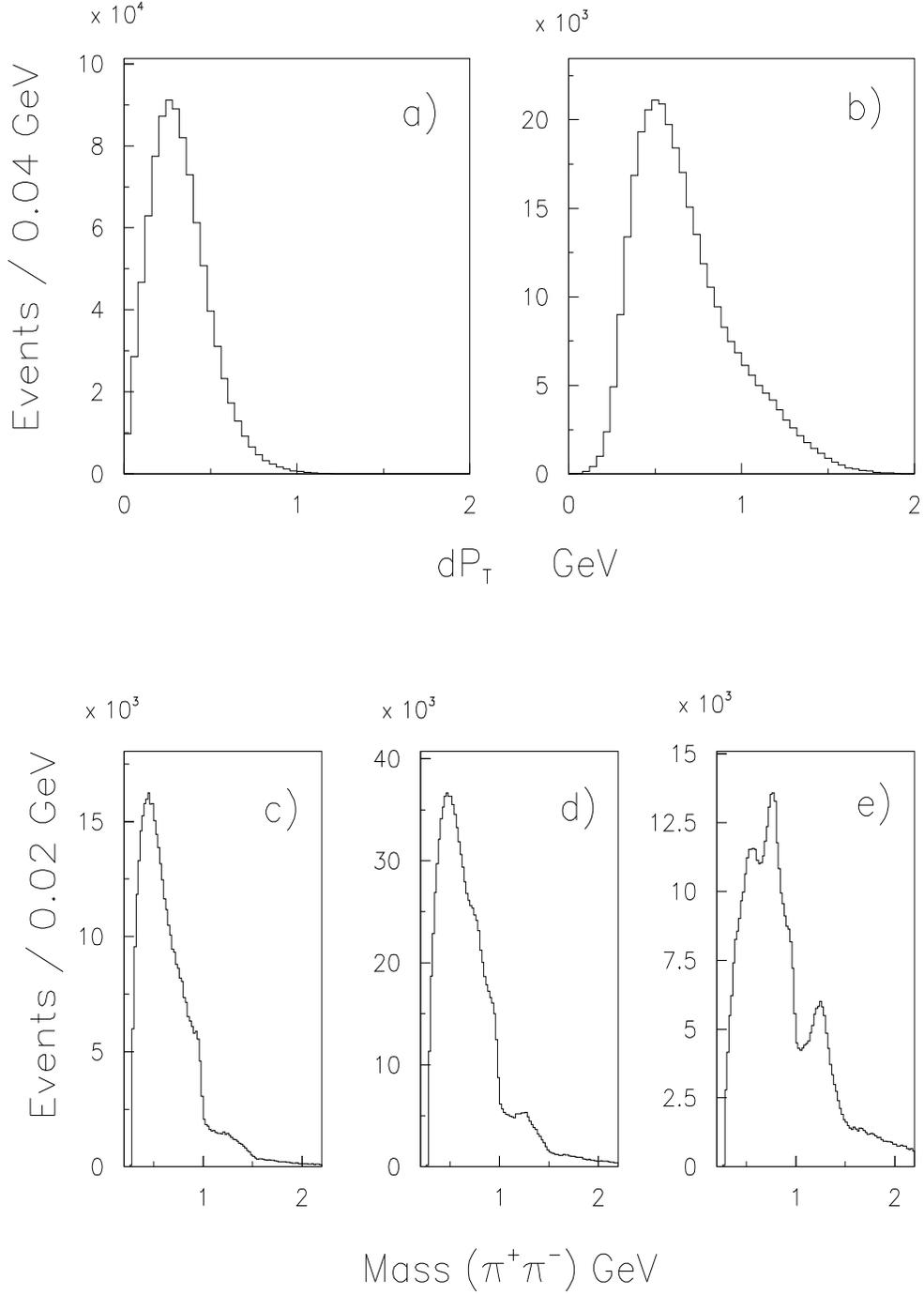,height=22cm,width=14cm}
\end{center}
\caption{$dP_T$ for a) LL and b) LR triggered events.
The \pipi mass spectrum for
c) $dP_T <   0.2$ GeV, d) $0.2 < dP_T <   0.5$ GeV and e) $dP_T >   0.5$ GeV.}
\label{fi:2pi}
\end{figure}

Figures~\ref{fi:2pi}a) and b) show the $dP_T$ spectrum for the
LL and LR trigger types respectively.
As can be seen the LL trigger type have access to events
with smaller $dP_T$.
It has been shown by Monte Carlo simulation that this effect is not
due to the fact that the LR events have additional trigger requirements
but it is due only to the fact that
the two protons recoil on the same (LL) or opposite (LR)
side of the beam direction~\cite{wa91corr}.
\par
The effect
that different cuts in $dP_T$ have on the $\pi^+\pi^-$ mass spectrum are shown
in
figures~\ref{fi:2pi}c), d) and e).
As can be seen, for $dP_T$~$<$~0.2~GeV there is effectively
no $\rho^0$(770) or \fmeson signals. These signals only become
apparent as $dP_T$ increases.
However the $f_0(980)$, which is responsible for the sharp drop in the
spectrum around 1~GeV, is clearly visible in the small
$dP_T$ sample.
\par
\begin{figure}[ht]
\begin{center}
\epsfig{file=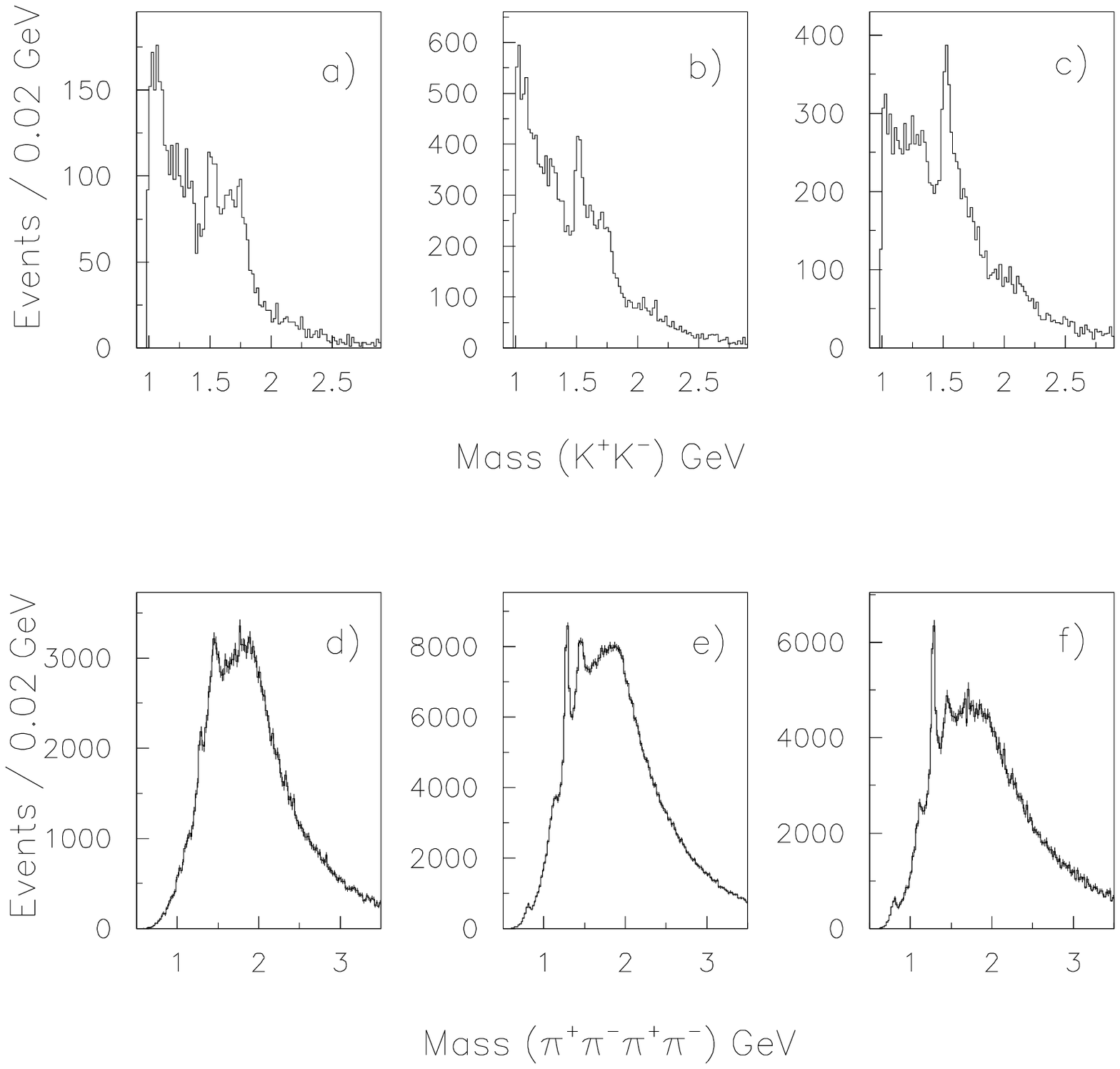,height=22cm,width=17cm}
\end{center}
\caption{$K^+K^-$ mass spectrum for a) $dP_T <   0.2$ GeV,
b) $0.2 <   dP_T <   0.5$ GeV and c) $dP_T >   0.5$ GeV and
the \pipipipi mass spectrum for d) $dP_T <   0.2$ GeV,
e) $0.2 <   dP_T <   0.5$ GeV and f) $dP_T >   0.5$ GeV.}
\label{fi:2k}
\end{figure}

Figures~\ref{fi:2k}a), b) and c) show the effect of the
$dP_T$ cut on the $K^+ K^-$ mass spectrum where
structures can be observed in the 1.5 and 1.7 GeV mass region which have
been previously identified as the
$f_{2}^\prime$(1525) and the $f_J(1710)$~\cite{re:WA76KK}.
As can be seen,
the $f_{2}^\prime$(1525) is produced dominantly at high $dP_T$,
whereas the $f_J(1710)$ is produced dominantly at low $dP_T$.
\par
In the \pipipipi  mass spectrum a dramatic effect is observed,
see figures~\ref{fi:2k}d), e) and f).
The $f_1(1285)$ signal has virtually disappeared at low $dP_T$
whereas
the $f_0(1500)$ and $f_2(1900)$ signals remain.
\par
A spin-parity analysis of the
$\pi^{+}\pi^{-}\pi^{+}\pi^{-}$
channel
has been performed~\cite{NEW4PI} using an isobar model~\cite{re:wa914pi}.
The $f_{1}(1285) $ is clearly seen in the
$J^{P}=1^{+}$~$\rho\rho $
and the $f_1(1285)$ signal almost disappears at
small $dP_T$.
In the $J^{P} =0^{+} \rho\rho $ distribution
a peak is observed at 1.45~GeV
together with a broad enhancement around 2~GeV.
The peak
in the $J^{P} =0^{+} \rho\rho $ wave
around 1.45 GeV remains
for $dP_T$~$\leq$~0.2~GeV while the
$J^{P} =0^{+} $
enhancement at 2.0~GeV becomes less
important: which shows that the $dP_T$ effect is not simply a $J^P$ filter.
\par
A fit has first been performed to the total
$J^{P} =0^{+} \rho\rho $
distribution using a K matrix formalism~\cite{KMATRIX}
including poles to describe the peak at 1.45 GeV as an interference between the
$f_0(1300)$, the $f_0(1500)$ together with a possible state at 2~GeV.
The resulting resonance parameters for
the $f_0(1300)$ and
$f_0(1500)$ are very similar to those found by Crystal Barrel~\cite{CBNEW}.
\par
The peak observed at 1.9~GeV, called the $f_2(1900)$,
is found to
decay to $a_{2}(1320)\pi$
and $f_{2}(1270)\pi\pi$ with $J^{PC}=2^{++}$.
At small $dP_T$ the $f_2(1900)$ signal is still important.
This is the first evidence of a non-zero spin resonance produced
at small $dP_T$ and hence shows that the $dP_T$ effect is not just
a $J^P$~=~$0^+$ filter.
\par
In addition to these waves,
a $J^{P} =2^{-}$ $a_{2}(1320)\pi $ wave was required in the fit.
The
$J^{P} =2^{-}$ $a_{2}(1320)\pi $ wave observed in this experiment is consistent
with the two $\eta_2$ resonances observed by Crystal Barrel~\cite{cbetapipi}
with both states decaying
to $a_2(1320) \pi $.
The $2^{-+} a_2(1320)\pi$ signal is
suppressed at small $dP_T$. This behaviour is
consistent with the signals being due to
standard $q \overline q$ states~\cite{closeak}.
\par
Similar effects are observed in all the other channels analysed to
date~\cite{NEWKKPI,NEW3PI}.
In fact it has been observed that
all the undisputed
$ q \overline q $ states
(i.e. $\rho^0(770)$, $\eta^{\prime}$, \fmeson, $f_1(1285)$,
$f_2^\prime(1525)$ etc.)
are suppressed as $dP_T$ goes to zero,
whereas the glueball candidates
$f_J(1710)$, $f_0(1500)$ and $f_2(1900)$ survive.
It is also interesting to note that the
enigmatic
$f_0(980)$,
a possible non-$q \overline q$ meson or $K \overline K$ molecule state does not
behave as a normal $q \overline q$ state.
\par
A Monte Carlo simulation of the trigger, detector acceptances
and reconstruction program
shows that there is very little difference in the acceptance as a function of
$dP_T$ in the different mass intervals considered
within a given channel and hence the
observed differences in resonance production can not be explained
as acceptance effects.
\par
\begin{figure}[htp]
\begin{center}
\epsfig{file=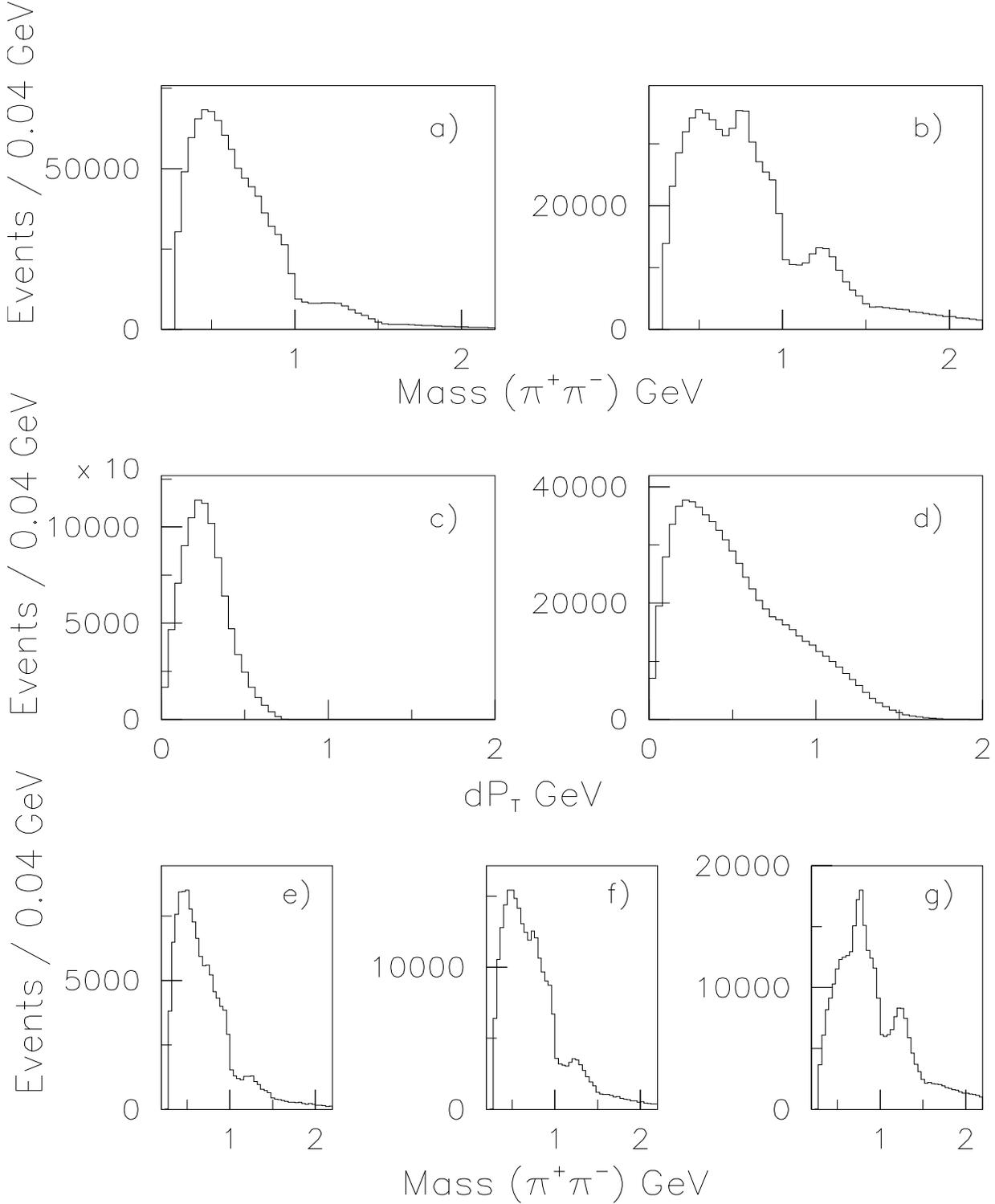,height=21cm,width=17cm}
\end{center}
\caption[tcut]{Results of cutting on the four momentum transferred at the
proton
vertices.
The $\pi^+\pi^-$ mass spectrum for a) $|t_f|$ $<$ 0.15 and $|t_s|$ $<$ 0.15
GeV$^2$ and
b) $|t_f|$ $>$ 0.15 and $|t_s|$ $>$ 0.15 GeV$^2$.
The $dP_T$ distribution for c) $|t_f|$ $<$ 0.15 and $|t_s|$ $<$ 0.15 GeV$^2$
and
d) $|t_f|$ $>$ 0.15 and $|t_s|$ $>$ 0.15 GeV$^2$.
The $\pi^+\pi^-$ mass spectrum for $|t_f|$ $>$ 0.15 and $|t_s|$ $>$ 0.15
GeV$^2$ and
e) $dP_T$ $<$ 0.2 GeV, f) 0.2 $<$ $dP_T$ $<$ 0.5 GeV and g) $dP_T$ $>$ 0.5 GeV.
}
\label{fi:tdep}
\end{figure}

It has previously been observed that the resonances produced in the central
region depend on the four momentum transferred from the fast ($t_f$) and slow
vertices ($t_s$)~\cite{wa76}.
The $\pi^+\pi^-$ mass spectrum is shown for the case where $|t_f|$ and $|t_s|$
are both less than 0.15 GeV$^2$ in figure~\ref{fi:tdep}a) and in
figure~\ref{fi:tdep}b) for the case when
$|t_f|$ and $|t_s|$ are both greater than 0.15 GeV$^2$. As can be seen
the amount of
$\rho^0(770)$ and $f_2(1270)$ does change as a function of this cut.
In figures~\ref{fi:tdep}c) and d) the $dP_T$ distribution for these two cases
is shown. As can be seen the events that have small $|t|$ are restricted
to small values of $dP_T$. To show that $dP_T$ is the
most important underlying dynamical effect the
$dP_T$ cut has been applied to the sample of events with large $|t|$.
Figures~\ref{fi:tdep}e), f) and g) show the events when
$|t_f|$ and $|t_s|$ are both greater than 0.15 GeV$^2$ for
$dP_T$~$\leq$~0.2~GeV,
0.2~$\leq$~$dP_T$~$\leq$~0.5~GeV and
$dP_T$~$\geq$~0.5~GeV respectively.
As can be seen the $dP_T$ cut still works in this sample.
\par
\section{Summary of the effects of the $dP_T$ filter}
\par
\begin{figure}[htp]
\begin{center}
\epsfig{file=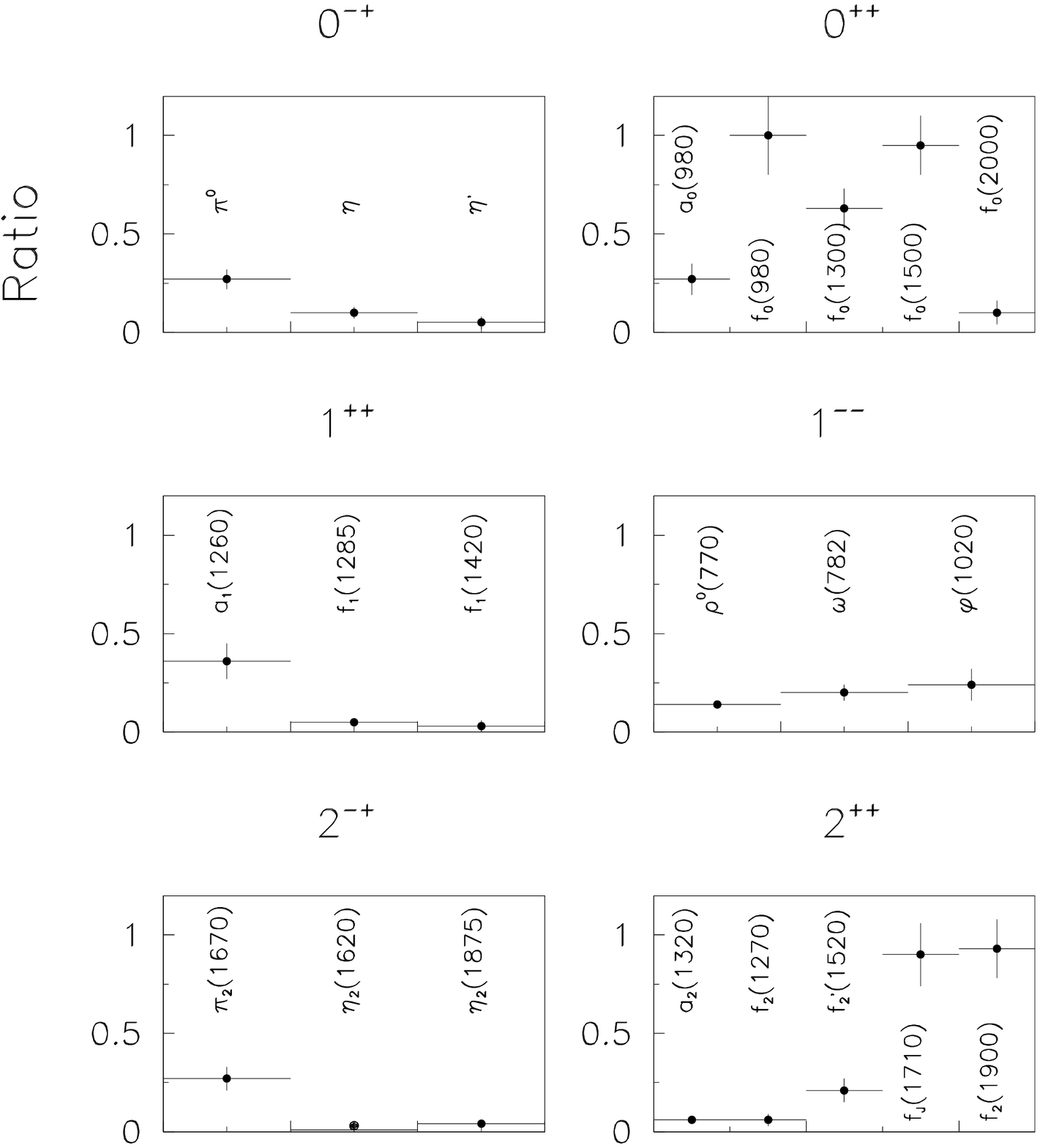,height=22cm,width=16cm}
\end{center}
\caption{The ratio of the amount of resonance with
$dP_T$~$\leq$~0.2 to the amount with
$dP_T$~$\geq$~0.5~GeV.
}
\label{fracratio}
\end{figure}

In order to calculate the contribution of each resonance as a function
of $dP_T$ the acceptance corrected mass spectra have been fitted with
the parameters of the resonances fixed to those obtained from the
fits to the total data. The results of these fits
are summarised in
table~\ref{frac} where the percentage of each resonance as a function of
$dP_T$ is presented.
Some of these values differ from previously published
values~\cite{NEW4PI,NEWKKPI}
due to an improved understanding and simulation of the experimental trigger.
Figure~\ref{fracratio} shows the ratio of the number of events
for $dP_T$ $<$ 0.2 GeV to
the number of events
for $dP_T$ $>$ 0.5 GeV for each resonance considered.
It can be observed that all the undisputed $q \overline q$ states
which can be produced in DPE, namely those with positive G parity and $I=0$,
have a very small value for this ratio ($\leq 0.1$).
Some of the states with $I=1$ or G parity negative,
which can not be produced by DPE,
have a slightly higher value ($\approx 0.25$).
However, all of these states are suppressed relative to the
interesting states, which could have a gluonic component, which have
a large value for this ratio.

\section{Conclusions}
\par
The results presented in this paper indicate the possibility of a
glueball-$q \overline q$ filter mechanism in central production.
All the
undisputed $q \overline q $ states are observed to be suppressed
at small $dP_T$, but the glueball candidates
$f_0(1500)$, $f_J(1710)$, and $f_2(1900)$ ,
together with the enigmatic $f_0(980)$,
survive.

\begin{table}[h]
\caption{Resonance production as a function of $dP_T$
expressed as a percentage of its total contribution.
The error quoted represents the statistical and systematic errors
summed in quadrature.}
\label{frac}
\begin{center}
\begin{tabular}{|c|c|c|c|c|} \hline
 & & & &  \\
$J^{PC}$ & Resonance&$dP_T$$\leq$0.2 GeV & 0.2$\leq$$dP_T$$\leq$0.5 GeV
&$dP_T$$\geq$0.5 GeV\\
 & & & & \\ \hline
 & & & & \\
$0^{-+}$
&$\pi^0$  &12 $\pm$ 2 & 44 $\pm$ 2 &44 $\pm$ 2 \\
& $\eta$  &6 $\pm$ 2  & 34 $\pm$ 2 &  60 $\pm$ 3 \\
& $\eta^\prime$  &3 $\pm$ 2 & 32 $\pm$ 2  &64 $\pm$ 3  \\
 & & & & \\ \hline
 & & & & \\
$0^{++}$
&$a_{0}(980)$  &14 $\pm$ 4  & 35 $\pm$ 4 &51 $\pm$ 7 \\
& $f_{0}(980)$  &22 $\pm$ 2 & 56 $\pm$ 3 &22 $\pm$ 3 \\
& $f_{0}(1300)$  &20 $\pm$ 2 & 48 $\pm$ 2 &32 $\pm$ 4 \\
& $f_{0}(1500)$  &23$\pm$ 2 & 53$\pm$ 3 &24 $\pm$ 4\\
& $f_{0}(2000)$  &5 $\pm$ 3 & 43 $\pm$ 5&52$\pm$ 5 \\
 & & & & \\ \hline
 & & & & \\
$1^{++}$
&$a_{1}(1260)$  &13 $\pm$ 3 & 51 $\pm$ 4 &36  $\pm$ 3 \\
& $f_{1}(1285)$  &3 $\pm$ 1& 35 $\pm$ 2 &61 $\pm$ 4 \\
& $f_{1}(1420)$  &2 $\pm$ 2 & 38 $\pm$ 2 &60 $\pm$ 4 \\
 & & & & \\ \hline
 & & & & \\
$1^{--}$
& $\rho(770)$  &8 $\pm$ 2 & 38 2 &54 $\pm$ 3 \\
& $\omega(782)$  &10 $\pm$ 2 & 40 $\pm$ 2 &49 $\pm $ 3 \\
& $\phi(1020) $  &10$\pm$ 3 & 48 $\pm$ 3 &42  $\pm$ 4\\
&  & & & \\ \hline
&  & & & \\
$2^{-+}$
& $\pi_{2}(1670)$  &11 $\pm$ 2 & 48 $\pm$ 4  &40 $\pm$ 4 \\
& $\eta_{2}(1620)$  &2 $\pm$ 1 & 42 $\pm$ 6 &54 $\pm$ 5 \\
& $\eta_{2}(1875)$  &1 $\pm$ 1  & 36 $\pm$ 7 &63 $\pm$ 7 \\
 & & & & \\ \hline
 & & & & \\
$2^{++}$
& $a_{2}(1320)$  &4 $\pm$ 4 & 35 $\pm$ 3 &61 $\pm$ 5 \\
& $f_{2}(1270)$  &4 $\pm$ 2 & 25 $\pm$ 2 &71  $\pm$ 3 \\
& $f_{2}^\prime(1520)$  &11 $\pm$ 3 & 37 $\pm$ 3 &52  $\pm$ 4 \\
& $f_{J}^\prime(1710)$  &26 $\pm$ 3 & 45 $\pm$ 2 &29  $\pm$ 4 \\
& $f_{2}(1900)$  &26 $\pm$ 2 & 46 $\pm$ 3 &28 $\pm$ 4 \\
 & & & & \\ \hline
\end{tabular}
\end{center}
\end{table}
\newpage
\newpage
\newpage
\newpage
\newpage
\clearpage


\newpage
\begin{thebibliography}{99}
\bibitem{re:lgt}
G. Bali  et al. (UKQCD),
Phys. Lett.  {\bf B309}  (1993) 378;

D. Weingarten, hep-lat/9608070;

J. Sexton et al., Phys. Rev. Lett. {\bf 75} (1995) 4563;

F.E. Close and M.J. Teper, ``On the lightest Scalar Glueball"
Rutherford Appleton Laboratory report no. RAL-96-040;
Oxford University report no. OUTP-96-35P


\bibitem{re:ISGURBNL}
N. Isgur, AIP. Conf. Proc. 185, Particles and Fields 36,
Glueballs, hybrids and exotic hadrons, (1988) 3
\bibitem{re:OZI}
S. Okuba, Phys. Lett.  {\bf 5} (1963) 165; \\
G. Zweig, CERN/TH 401,402,412 (1964); \\
J. Iizuka, Prog. Theor. Phys. Suppl.  {\bf 37-38} (1966) 21.
\bibitem{closerev} D.Robson, Nucl Phys {\bf B130} (1977) 328; \\
F.E. Close, Rep. Prog. Phys. {\bf 51} (1988) 833.
\bibitem{re:expt}
T.A. Armstrong {\em et al.,} Nucl. Instr. and Methods {\bf A274} \rm (1989)
165;\\
F. Antinori {\em et al.,} Il Nuovo Cimento {\bf A107 } (1994) 1857.
\bibitem{pred} S.N. Ganguli and D.P. Roy, Phys. Rep. {\bf 67} (1980) 203.
\bibitem{wa76} T.A.\thinspace Armstrong et al., Zeit. Phys. {\bf C 51} (1991)
351.
\bibitem{closeak}
F.E. Close and A. Kirk, Phys. Lett. {\bf B397 } \rm (1997) 333.
\bibitem{wa91corr} D.\thinspace Barberis et al., Phys. Lett. {\bf B388} (1996)
853.
\bibitem{re:WA76KK} T.A.\thinspace Armstrong et al., Phys. Lett. {\bf B227}
(1989) 186.
\bibitem{NEW4PI}
D. Barberis {\em et al.,} Phys. Lett. {\bf B413} \rm (1997) 217.
\bibitem{re:wa914pi}
S. Abatzis {\em et al.,} Phys. Lett. {\bf B324 } \rm (1994) 509.
\bibitem{KMATRIX}
S.U. Chung {\em et al.,} Ann. d. Physik. {\bf 4} (1995) 404.
\bibitem{CBNEW}
A. Abele {\em et al.,} Nucl. Phys. {\bf A609} (1996) 562.
\bibitem{cbetapipi}
C. Amsler {\em et al.,} Zeit. Phys. {\bf C71 } (1996) 227.
\bibitem{NEWKKPI}
D. Barberis {\em et al.,} Phys. Lett. {\bf B413} \rm (1997) 225.
\bibitem{NEW3PI}
D. Barberis {\em et al.,} hep-ex/9801003, Submitted to Phys. Lett.
\end{thebibliography}
\end{document}